\documentclass[twocolumn]{aastex62}
\usepackage{multirow}
\usepackage{makecell}
\usepackage{amsmath}
\usepackage{scrextend}
\usepackage{tablefootnote}

\begin{document}

\title{A Tight Relation between Spiral Arm Pitch Angle and Protoplanetary Disk Mass}

\author[0000-0002-3462-4175]{Si-Yue Yu}
\affiliation{Kavli Institute for Astronomy and Astrophysics, Peking University, Beijing 100871, China}
\affiliation{Department of Astronomy, School of Physics, Peking University, Beijing 100871, China}

\author[0000-0001-6947-5846]{Luis C. Ho}
\affiliation{Kavli Institute for Astronomy and Astrophysics, Peking University, Beijing 100871, China}
\affiliation{Department of Astronomy, School of Physics, Peking University, Beijing 100871, China}

\author[0000-0003-3616-6822]{Zhaohuan Zhu}
\affiliation{Department of Physics and Astronomy, University of Nevada, Las Vegas, 
4505 South Maryland Parkway, Las Vegas, NV 89154, USA}

\begin{abstract}
We use two-dimensional Fourier transformation to measure the pitch angle ($\varphi$) 
of the dominant spiral Fourier mode 
of well-defined spiral arms in 13 protoplanetary disks, 
making use of near-infrared scattered-light images of AB~Aur, SAO~206462, 
MWC~758, V1247~Ori, HD~142527, DZ~Cha, LkH$\alpha$~330, and HD~100453, and 
ALMA millimeter continuum images of Elias~2-27, IM~Lup, AS~205, and HT~Lup.  
We find that the measured pitch angle correlates strongly with disk mass ($M_{D}$),
such that more massive protoplanetary disks have smaller pitch angles, following
$|\varphi| = -(7.8\pm1.7)\log(M_{D}/M_{\odot})+(2.7\pm2.6)$.  Interestingly, 
four disks with a known companion (HD~142527, HD~100453, AS~205, and HT~Lup) share the 
same trend.  Such a strong dependence of spiral arm pitch angle on disk mass 
suggests that the disk mass, independent of the formation mechanism, plays 
a fundamental role in 
determining the arm tightness of the observed spiral structure.
The physical origin of the $\varphi-M_D$ relation is still not clear.
The pitch angle of spiral arms in 
protoplanetary disks provides an independent constraint on the disk mass.
\end{abstract}


\section{Introduction}

Well-defined spiral structure has been detected in protoplanetary disks owing 
to high-spatial resolution observations, in both near-infrared (NIR) 
scattered-light images of AB~Aur \citep{Hashimoto2011}, SAO~206462 
\citep{Muto2012, Garufi2013}, MWC~758 \citep{Grady2013, Benisty2015}, V1247~Ori
\citep{Ohta2016}, HD~142527 \citep{Avenhaus2017}, HD~100453 \citep{Wagner2015, 
Benisty2017}, DZ~Cha \citep{Canovas2018}, and LkH$\alpha$~330 \citep{Uyama2018},
and in Atacama Large Millimeter/submillimeter Array (ALMA) millimeter continuum
images of Elias~2-27 \citep{Perez2016, DSHARP3}, IM~Lup \citep{DSHARP3}, 
WaOph~6 \citep{DSHARP3}, AS~205 \citep{DSHARP4}, and HT~Lup \citep{DSHARP4}.
Spiral arms in HD~100546 exhibit different chirality \citep{Follette2017} and 
thus are not well-defined.  Because of the large dust scattering opacity, NIR 
scattered-light observations detect structure on the disk surface, while ALMA 
millimeter continuum observations probe the cold dust in the disk midplane.
The spirals of MWC~758 coexist both in the NIR and millimeter continuum images 
\citep{Boehler2018, Dong2018a}, but the latter are much more asymmetric.  Most of the 
disks with spiral arms observed in the NIR show a peculiar dip in the infrared 
spectral energy distribution that may indicate a lack of warm dust near the 
central star \citep{Strom1989, Skrutskie1990, Garufi2018}, suggesting that the mechanism 
to form gaps may be related to spiral arm formation.  These results, however, may 
be affected by small sample size or observational selection effects.  

Pitch angle ($\varphi$), defined as the angle between the tangent of a spiral 
arm and the azimuthal direction, describes the degree of tightness of the arm.
The classic quasi-stationary density wave theory, proposed by \cite{LinShu64}, 
is perhaps the most successful framework to explain spiral structure in 
galaxies.  In this framework, a number of works aimed to understand linear 
nonaxisymmetric density perturbations, including dipole or spiral perturbations,
in gaseous and collisionless self-gravitating disks \citep{Adams1989, Shu1990, 
Noh1991, Laughlin1996}.  However, these linear stability analyses have not 
investigated the pitch angle of spiral arms in the disks.  In contrast, 
\cite{Rafikov2002} studied the nonlinear propagation of a one-armed spiral wake
launched by a planet embedded in a disk.  Using weakly nonlinear density wave 
theory in the WKB limit, Rafikov proposed that the pitch angle of 
planet-generated spiral arms depends on the sound speed of the disk and the 
location of the planet.  \cite{Muto2012} and \cite{Benisty2015} applied this 
scenario to SAO~206462 and  MWC~758, respectively, to infer the properties of 
their protoplanetary disks and the position of the hypothetical unseen planet.  
The recent studies of \cite{BaeZhu2018} and \cite{Miranda2019} further show, based on linear theory, 
that in addition to this main arm, a secondary arm can arise in the inner part 
of the disk.  Similarly, in hydrodynamical simulations the presence of a 
massive companion can  induce spiral arms \citep[e.g.][]{Kley2012, Zhu2015} 
that well match observations \citep{Dong2015a, Dong2016}. \cite{Zhu2015}, in 
particular, showed that, in addition to the sound speed in the disk and the 
location of the perturber, the pitch angle of planet-induced spiral arms also 
depends on the mass of the planet, such that arms become more open with a more 
massive perturber.

Gravitational instability, often explored using simulations \citep[e.g.,][]{
Lodato2004, Rice2004, Forgan2011}, is another possible mechanism to generate 
spiral arms in circumstellar disks.  One apparent shortcoming of these 
simulations is that they tend to produce a significantly larger number of arms 
than the two normally observed.  Recent studies show that such simulations of 
gravitationally unstable disks are also able to generate two-armed spirals 
\citep{Dong2015b, Tomida2017}, which qualitatively match the observed arms in 
the Elias~2-27 disk \citep{Meru2017, Tomida2017}.  But these simulated spirals disappear 
in a few rotations \citep{Tomida2017}, implying that their shape and, hence, 
the pitch angle of the simulated spirals also change frequently with time.

Pressure variations due to shadowing from a misaligned inner disk have been 
proposed to trigger spiral arms observed in scattered light \citep{Montesinos2016, Montesinos2018}. 
As the simulations of \cite{Montesinos2016} show, a less massive disk may have 
more open arms, which, however, would eventually evolved into tight arms.

Note that these mechanisms of spiral arm formation are not necessarily mutually
exclusive. For example, a tidal interaction theoretically can induce an 
external perturbation, which results in spiral structure obeying density 
wave theory, with, perhaps, gravitational instability participating in it, 
making the structure more complicated.

The pitch angle of spiral arms may shed light on their formation mechanism. 
We aim to establish the dependence of pitch angle on the properties of 
protoplanetary disks to probe the physical origin of spiral arms.

\section{Data}

This study makes use of the following data: \\
VLT/SPHERE images of MWC~758 \citep{Benisty2015}, HD~142527 
\citep{Avenhaus2017}, DZ~Cha \citep{Canovas2018}, and HD~100453 
\citep{Benisty2017}; VLT/NACO image of SAO~206462 \citep{Garufi2013}; 
Subaru/HiCIAO images of AB~Aur \citep{Hashimoto2011}, LkH$\alpha$~330 
\citep{Uyama2018}, and V1247~Ori \citep{Ohta2016}, and ALMA millimeter 
continuum images of Elias~2-27, IM~Lup, WaOph~6, AS~205, and HT~Lup from the 
Disk Substructures at High Angular Resolution Project (DSHARP) \citep{DSHARP1, 
DSHARP3, DSHARP4}.

With the exception of Elias~2-27, WaOph~6, AS~205, and HT~Lup, the masses of 
the protoplanetary disks ($M_{D}$) are from \cite{Dong2018b}, who converted 
the dust submillimeter continuum emission at 880\,$\micron$ to total mass assuming a dust opacity 
of $\kappa$\,=\,3\,cm$^2$\,g$^{-1}$ and a gas-to-dust mass 
ratio of 100.  We compute $M_D$ for Elias~2-27, WaOph~6, AS~205, and HT~Lup 
following the procedure described in \cite{Dong2018b}, adopting the disk radius  
in our Table~1 and spectral energy distributions (SED) and other parameters collected by \cite{DSHARP1}. 
The uncertainties of the disk masses are estimated by assigning fractional errors of $15\%$, $15\%$, and $30\%$
to the submillimeter fluxes, dust opacity, and gas-to-dust mass ratio, respectively.

To roughly estimate Toomre's (1964) $Q$ of the disk, we assume a Keplerian 
disk heated by the irradiation of a central star. The dust temperature follows 
$T=\{\phi L_*/8\pi\sigma_{\rm SB} r^2\}^{1/4}$, where $\phi$, set to 0.02 for 
simplicity, is the flaring angle, $L_*$ is the luminosity of the central star, 
and $\sigma_{\rm SB}$ is the Stefan-Boltzmann constant.  Then we have sound speed 
$c_s=\sqrt{k_{\rm B} T/2.3\,m_p}$, with the Boltzmann constant $k_{\rm B}$ 
and proton mass $m_p$.  The mean surface density is estimated 
as $\overline{\Sigma}=M_D/\pi R_D^2$, for disk radius $R_D$.  The mean $Q$ is 
derived as $Q=c_s\kappa/\pi G \overline{\Sigma}$, where $c_s$ and epicyclic 
frequency $\kappa$ are estimated at $r=R_D$. The disk aspect ratio $H/R$, with 
$H\equiv c_s/\Omega$ and $\Omega$ the angular velocity, is evaluated at 
$r=R_D$.  

Table~1 lists the parameters for the protoplanetary disks and their central 
stars used in this work: inclination angle ($i$), position angle (PA), 
disk radius ($R_D$), mass of central star ($M_*$), flux density at 880\,$\mu$m ($f_{880\mu \rm m}$), 
disk mass ($M_{D}$), luminosity of central star ($L_*$), mean Toomre's $Q$, disk aspect ratio ($H/R$), 
and pitch angle ($\varphi$).

\section{Measurement of Pitch Angle}

\begin{figure*}
\figurenum{1}
\centering
\plotone{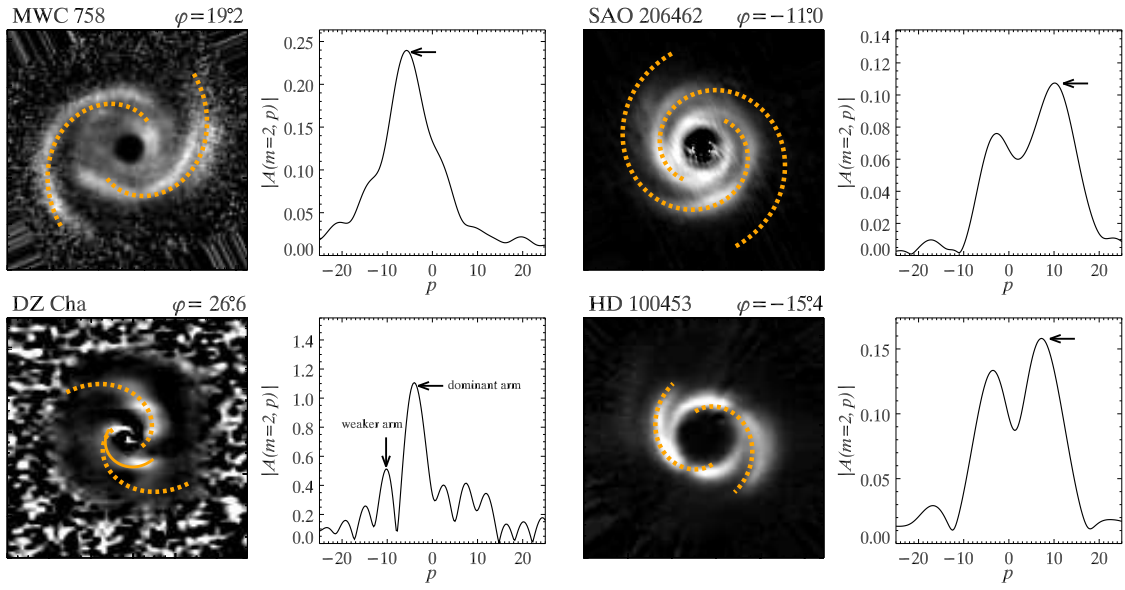}
\caption{Illustration of pitch angle measurement for MWC~758, SAO~206462, 
DZ~Cha, and HD~100453. (Left) Deprojected $r^2$-scaled NIR scattered-light 
images, overploted with synthetic arms, marked by dotted curve, with measured pitch 
angle of dominant spiral Fourier mode.  (Right) Amplitude of Fourier spectra ($|A(m,p)|$), with arrow 
indicating the peak selected to calculate pitch angle. 
For DZ~Cha, the peaks corresponding to the dominant open arm and the weaker, secondary tight arm (solid curve on the left) are labeled.
}
\end{figure*}

\begin{figure*}
\figurenum{2}
\centering
\plotone{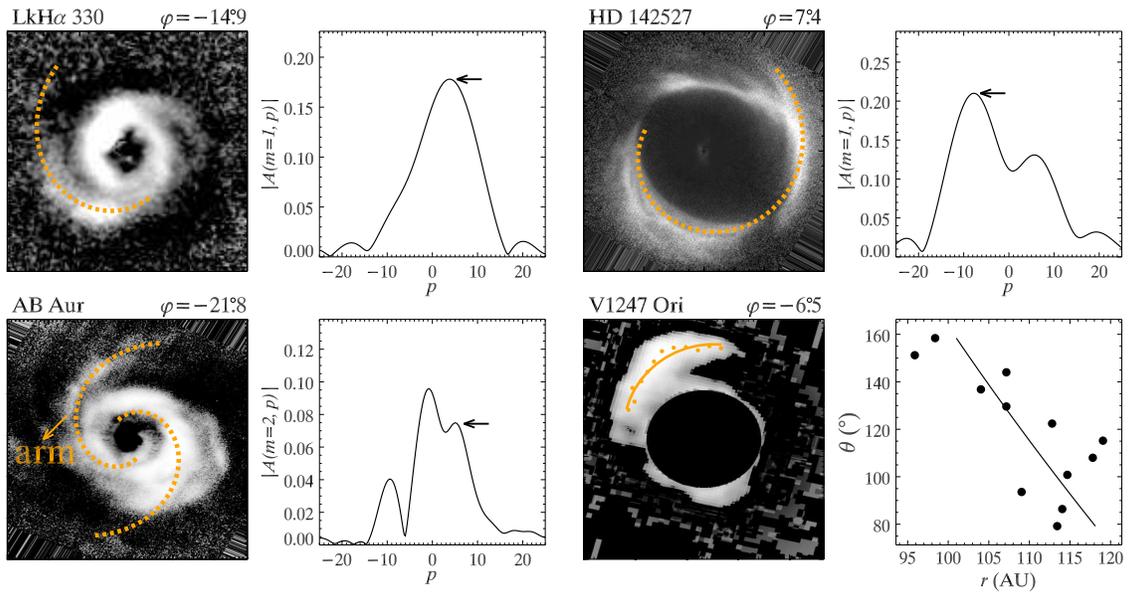}
\caption{Same as in Figure~1, but for LkH$\alpha$~330, HD~142527, 
and AB~Aur. For V1247~Ori, the pitch angle is measured by 
identifying the local maxima (orange points in left panel; black 
points in right panel) along the arm, with an azimuthal step of $7\degr$,
and then fitting a logarithmic function to the spiral positions.
The solid line marks the best-fit logarithmic function.}
\end{figure*}

\begin{figure*}
\figurenum{3}
\centering
\plotone{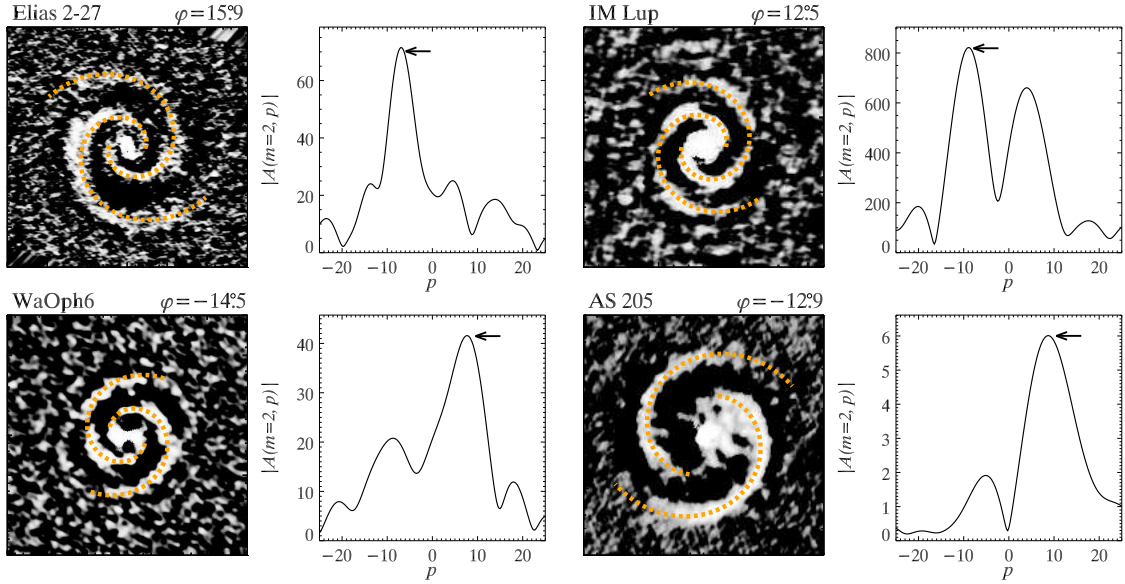}
\caption{Same as in Figure~1, but for the residual non-axisymmetric component 
of deprojected ALMA millimeter images of Elias~2-27, IM~Lup, WaOph~6, and 
AS~205N.}
\end{figure*}

\begin{figure*}
\figurenum{4}
\centering
\includegraphics[width=18cm]{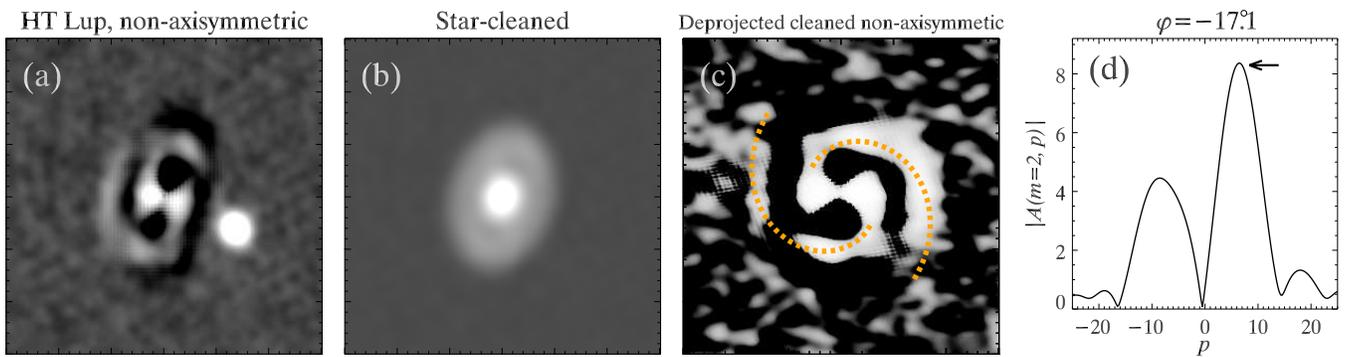}
\caption{Spiral structure and pitch angle measurement for HT~Lup: 
(a) residual non-axisymmetric component of ALMA millimeter image;
(b) star-cleaned image;
(c) residual non-axisymmetric component of deprojected star-cleaned image;
(d) amplitude of Fourier spectra ($|A(m,p)|$) for residual deprojected 
star-cleaned image, with arrow indicating the peak selected to calculate pitch 
angle. 
}
\end{figure*}

For the NIR scattered-light images, we scale each pixel with the square of its 
distance from the star ($r^2$) to compensate for the $r^{-2}$ dependence of 
the stellar illumination.  The arms in the five ALMA millimeter continuum 
images are relatively fainter. To reduce the potentially adverse effect of 
radial variation of intensity, similar to the strategy in \cite{DSHARP3}, we 
construct the axisymmetric component of the disk by finding the median 
intensity within annuli of 1\,AU width, with $i$ and PA fixed to the values of
the disk, and then subtract the axisymmetric component from the image to obtain
the residual non-axisymmetric component for the pitch angle measurement.

We use $i$ and PA to deproject the image from the previous step to its face-on 
orientation, employing the {\tt IRAF} task {\tt geotran}.  As the ratio of 
scale height to radius varies from $\sim 10^{-2}$ near the star/disk interface 
to $\sim 10^{-1}$ near the outer edge of the disk \citep[see][]{Kenyon1987, 
Dullemond2002}, the disk is geometrically thin.  Most of the disks in our 
sample are relatively face-on, with $i \lesssim40\degr$, mitigating projection 
effects.

Two-dimensional discrete Fourier transformation (2DDFT) is a well-defined and 
the most widely used technique to measure the pitch angle of spiral arms in disk galaxies 
\citep[e.g.,][] {Kalnajs1975, Iye1982, Krakow1982, Puerari1992, Puerari1993, Block1999, Davis2012, Yu2018a}.   
The 2DDFT method decomposes images into Fourier components of different
radial and azimuthal frequencies, i.e. spirals of different pitch angles and number of arms, and then chooses the dominant Fourier mode to calculate the pitch angle. In the context of galactic disks, this methodology has been useful in identifying physical relationships between spiral arm pitch angle and the global structure, mass, and kinematics of galaxies \citep{Ma2002, Seigar2005, 
Seigar2006, Seigar2008, Savchenko2013, Kendall2015, Yu2018a, Yu2018b, Yu2019}.
This work uses the 2DDFT method discussed in 
detail by \cite{Yu2018a}.  Here, we just briefly summarize a few essential 
points.  We transform the deprojected images into polar coordinates and 
decompose the light distribution into a superposition of 2D Fourier components 

\begin{eqnarray}
\begin{split}
A(m, p) = \frac{1}{D} \int ^{\text{ln}(r_{\text{out}})}_{\text{ln}(r_{\text{in}})} \int^{\pi}_{-\pi} \\
\sum_{j=1}^{N} I_j(r_j, \theta_j) \delta(\mu-\mu_j) \delta(\theta-\theta_j) e^{-i(m\theta + p \mu)} d\theta d\mu,
\end{split}
\end{eqnarray}

\noindent
with normalization factor $D =  \sum_{j=1}^N I_j$, where $I_j$ is the intensity
of the $j$th pixel at $(r_j, \theta_j)$, $r_{\text{in}}$ and $r_{\text{out}}$ 
the inner and outer boundary of the spiral structure, $N$ the number of pixels 
within the radial range, and $\mu \equiv {\rm ln}\, r$.  
The most prominent peak $p_{\rm max}$ of the power spectrum of spiral Fourier mode 
$m$ is identified to calculate the pitch angle of the dominant spiral Fourier mode:
$\varphi = \arctan{\left(- \frac{m}{p_{\rm max}} \right)}$.  
As the 2DDFT method uses all pixel values within the imposed radial range to calculate the Fourier
spectrum, the resulting pitch angle is an average value, with the flux as weighting, of different arms over the imposed radial range. In the event that any given arm is strong enough to dominate the Fourier spectrum, the resulting pitch angle will only trace such an arm.
Following the strategy of \cite{Yu2018a}, the uncertainty of $\varphi$ is estimated by 
repeating the measurement over three radial ranges: [$r_{\rm in}$, $r_{\rm out}-\Delta r$], [$r_{\rm in}+\Delta r$, $r_{\rm out}$], and [$r_{\rm in}+\Delta r/2$, $r_{\rm out}-\Delta r/2$], where $\Delta r=(r_{\rm out}-r_{\rm in})*20\%$, 
to account for both the uncertainty of manually choosing 
the spiral arm boundary and the radial variation of pitch angle. 

The 2D Fourier spectra ($|A(m,p)|$) and the synthetic arms with measured 
$\varphi$ are presented in Figures 1--4.  As shown in Figure~1, the spiral arms
in the NIR scattered-light images of MWC~758, SAO~206462, DZ~Cha, and 
HD~100453 clearly have two arms, and their Fourier spectra are dominated by the
$m=2$ mode.  The prominent peak is selected to calculate $\varphi$.  
For MWC~758, we give a pitch angle of $19\fdg2\pm1\fdg0$. There are deviations between the observed spirals and the synthetic spirals, owing to the slight asymmetry of the two main arms and other
short arms associated with the end of the right main arm.
The 2DDFT method measures the average pitch angle for them. 
Our measured pitch angle is significantly larger than the result from Dong et al. 
(\citeyear{Dong2015a}; $\sim$\,11$\degr$), who determined the pitch angle by 
identifying the location of the spiral arms.
Compared with their method, the 2DDFT technique has two major advantages.
First, 2DDFT simultaneously considers all the fine spiral structures.  
\cite{Benisty2015} showed that, apart from the two main arms, the disk of 
MWC~758 has four additional non-axisymmetric features.  Moreover, 2DDFT 
naturally weights by the intensity when calculating the Fourier components, 
making it sensitive to the structural information of the dominant spiral Fourier modes.
This is also the reason why the measured pitch angle for DZ~Cha, 
$\varphi=26\fdg6\pm3\fdg1$, can only well trace the strong open arm, which dominates
the Fourier spectrum with a single narrow peak.
Multiple density waves may exist in DZ~Cha. The dominant one is perhaps more closely associated
with the formation physics of the global spiral structure.
As discussed in Section 4, the pitch angle of the dominant component correlates strongly with 
the physical properties of the system.  
Nevertheless, the weaker arm, from visual inspection, should have a much smaller pitch angle, and it is expected to contribute another peak to the left of the dominant peak. We thus use the secondary peak of the spectrum, as indicated
in Figure 1, to calculate the pitch angle of the weaker arm, resulting in $10\fdg3$.
The mean pitch angle of the two arms of DZ~Cha is $18\degr$. The impact of 
DZ~Cha on our results are discussed in Section 4.
The measured pitch angle of SAO~206462 
($11\fdg8\pm1\fdg7$) is consistent with that reported in Dong et al. (\citeyear{Dong2015a}; 
$\sim$\,11$\degr$).  The arms of HD~100453 are not perfectly symmetric, and 
the measured pitch angle, $\varphi=15\fdg4\pm0\fdg7$, traces their average 
tightness.

Figure~2 presents the results for the sources without two clear arms in NIR 
scattered-light images.  The Fourier spectrum of LkH$\alpha$~330 is dominated 
by the $m=1$ mode of the stronger of its two arms; the resulting pitch angle 
($14\fdg9\pm2\fdg8$) is entirely consistent with the results of 
\cite{Uyama2018}, who quoted $\sim$\,12$\degr$ for the strong arm and 
$\sim$\,16$\degr$ for the weaker one.  For AB~Aur, as the most prominent peak of 
the $m$\,=\,$2$ Fourier mode 
has $p\approx0$, the resulting ``pitch angle'' of $\sim90\degr$ corresponds to the central ring, which is not so symmetric in shape and light distribution.
We use, instead, the secondary peak of the Fourier spectrum to 
calculate the pitch angle ($\varphi=21\fdg8\pm1\fdg3$), whose synthetic arms well trace the main spiral arm 
indicated by the arrow in Figure 2.
While AB~Aur exhibits
other arm pieces \citep{Hashimoto2011} and 
small-scale gaseous arms inside the dust cavity \citep{Tang2017}, these short,
faint arm pieces are likely a result of local instabilities and differential motion, stemming from
formation physics very different from that of global spirals. 
The global disk
properties are not expected to significant affect such small-scale inner spirals. As we aim to systematically
investigate the dependence of the global spiral pitch angle on disk properties, we do not consider the
small-scale inner spirals. But, interestingly, the pitch angle of the inner western spirals ($\sim20\degr$)
reported by \cite{Tang2017} is consistent with our measured pitch angle for the global spirals.
HD~142527 has many small-scale feathery 
arms, but its Fourier spectrum is dominated by the $m$\,=\,1 mode, which 
results in a tightly wound spiral of $\varphi=7\fdg4\pm0\fdg5$.  The 2DDFT 
method fails to measure the pitch angle for V1247~Ori, since its spiral arm is 
too tightly wound and too short in radial extent.  We measure its pitch angle 
by identifying a number of local maxima within the arm, in azimuthal steps of 
$7\degr$. Then the pitch angle is estimated by fitting a logarithmic function 
to the positions of the local maxima in the arm.

Figures~3--4 plot the pitch angle measurements for the five ALMA continuum 
images.  Elias~2-27, IM~Lup, WaOph~6, and AS~205 have two symmetric arms with 
Fourier spectra dominated by an $m=2$ mode. We find pitch angle 
$\varphi=15\fdg9\pm2\fdg4$ for Elias~2-27, $\varphi=14\fdg5\pm3\fdg1$ for 
WaOph~6, and $\varphi=12\fdg9\pm1\fdg1$ for AS~205; these values are consistent
with those reported in \cite{DSHARP3} and \cite{DSHARP4}.  We assign a global 
pitch angle $\varphi=12\fdg5\pm2\fdg7$ to IM~Lup; the pitch angle of the 
spiral arms in this object decreases from $\sim19\degr$ in the inner region to 
$\sim10\degr$ in the outer part \citep{DSHARP3}. 

As shown in Figure~4a, the spiral arms in the non-axisymmetric component of 
HT~Lup are not symmetric.  In particular, the clear spiral arm to the east 
was not identified by \cite{DSHARP4} for measuring pitch angle, whereas the
corresponding arm to the west is nearly invisible.  Note that there is a 
strong central bar in this system.  A close stellar companion to the southwest 
may potentially contaminate the Fourier spectra.  We removed the star by 
fitting a Gaussian function to it in the residual non-axisymmetric component 
image, and then subtracting it from the original image to construct the 
star-cleaned image (Figure~4b).  We then generate a residual star-cleaned 
non-axisymmetric component image and deproject it (Figure~4c) for Fourier 
decomposition.  The Fourier spectrum (Figure~4d) presents a prominent peak, 
which yields $\varphi=17\fdg1\pm0\fdg8$.  Our measured pitch angle is 
significantly larger than that in \cite{DSHARP4}, probably due to their  
omission of the eastern arm and 
the arm intensity varying significantly with radius.

\begin{deluxetable*}{ccccccccccccc}
\def\a{\hskip 1.5 mm}
\def\b{\hskip -1.5 mm}
\def\c{\hskip  1.3 mm}

\tabcolsep 8 pt
\tablenum{1}
\tablecaption{Spiral Arm Pitch Angles and Properties of Protoplanetary Disks and their Central Stars}
\tablehead{
\colhead{Object} &
\colhead{$i$} &
\colhead{PA} &
\colhead{$R_{D}$} &
\colhead{$M_*$} &
\colhead{$f_{880\mu \rm m}$} &
\colhead{$M_{D}$} &
\colhead{$L_{*}$} &
\colhead{$Q$} &
\colhead{$H/R$} &
\colhead{$|\varphi|$} &
\colhead{References} \\
\colhead{} &
\colhead{(deg)} &
\colhead{(deg)} &
\colhead{(AU)} &
\colhead{($M_{\odot}$)} &
\colhead{(mJy)} &
\colhead{(0.01 $M_{\odot}$)} &
\colhead{($L_{\odot}$)} &
\colhead{} &
\colhead{} &
\colhead{(deg)} \\
\colhead{(1)} &
\colhead{(2)} &
\colhead{(3)} &
\colhead{(4)} &
\colhead{(5)} &
\colhead{(6)} &
\colhead{(7)} &
\colhead{(8)} &
\colhead{(9)} &
\colhead{(10)} & 
\colhead{(11)} &
\colhead{(12)}
}
\startdata
MWC~758              & 21    &  65  &151      &1.68  &180    & 1.18           &\a8.5&      11   & 0.08&$19.2\pm 1.0$       &1,2,3,2,3,2\\
SAO~206462         & 11.5 &  64  &156      &1.70   &620   & 4.08           &\a8.8&      3    &0.08&$11.0\pm 1.0$       &4, 2,3,2,3,5\\
LkH$\alpha$~330  &  31   & 91   &170      &2.12   &210   & 3.39           &12.8&        5    &0.08&$14.9\pm 2.8$      &6,2,3,2,3,7\\
DZ~Cha                 & 43   & 176  &22      &0.51     &21     & 0.20           &\a0.6&       16  &0.06&$26.6\pm 3.1$       &8,$\cdots$,3,8,3,8\\
AB~Aur                  & 36.6 &26.8 &230       &2.50  &317   & 1.50           &43.8&        14  &0.09&$21.8\pm 1.3$      &9,10,3,11,3,11\\
HD~142527           &  20  &299   &300       &1.70  &3310  &33.8          &\a9.9&        0.5 &0.10&$ \a7.4\pm 0.5$    &12,13,3,13,3,5\\
V1247~Ori             & 31.3 &104  &190       &1.91  &292   & 7.64          &15.8&         2   &0.09&$ \a6.5\pm 0.7$    &14,15,3,15,3,5\\
HD~100453           &38  &142       & 48   &1.53     &464   &1.74            &6&             5    &0.06&$ 15.4\pm 0.7$    &16,16,3,17,3,5\\
Elias~2-27             & 56.2  &118.8 &300    &\b0.5 &666    &10.4           &1.0&          0.6 &0.13&$ 15.9\pm 2.4$    &18,19,20,$\cdots$,$\cdots$,21\\
IM~Lup                  &47.5 & 144.5 & 300  & \b0.6 &582    &18.4           & 0.9&           0.4  &0.12&$12.5\pm2.7$    &18,22,3,22,3,22         \\
WaOph~6              &47.3  & 174.2 &137  &\b0.7  &386    &2.17          & 2.9 &           3  &0.10&$14.5\pm2.3$       &18,$\cdots$,20,$\cdots$,$\cdots$,21       \\
AS~205               &20.1&114.0&60&   \b0.9          &872    &3.28          &2.1&              2   &0.07&  $12.9\pm1.1$    &23,$\cdots$,20,$\cdots$,$\cdots$,24    \\
HT~Lup               &48.1&166.1&37&   \b1.7         &175     &0.53          &5.5&              17   &0.05&  $17.1\pm0.8$   &23,$\cdots$,20,$\cdots$,$\cdots$,25  \\
\enddata
\tablecomments{
Col. (1): Source name.
Col. (2): Inclination angle.
Col. (3): Position angle; references same as in Col. (2).
Col. (4): Radius of the disk in millimeter continuum emission, except for 
DZ~Cha, whose disk radius is estimated using NIR scattered-light images from \cite{Canovas2018}; disk radius of WaOph~6, AS~205, and HT~Lup
are determined in this work using ALMA millimeter images.
Col. (5): Mass of central star.
Col. (6): Flux density at 880\,$\mu$m ($f_{880\mu \rm m}$); $f_{880\mu \rm m}$ of Elias~2-27, WaOph~6, AS~205, and HT~Lup are obtained by interpolation of SED collected by \cite{DSHARP1}; 
others are determined by converting the flux densities given in the reference listed in Col. (12) 
to flux densities at 880\,$\mu$m through a power law $f_{\nu}\propto\nu^{2.4}$.
Col. (7): Disk mass; disk masses of Elias~2-27, WaOph~6, AS~205, and HT~Lup are derived in this work following the procedure in \cite{Dong2018b}.
Col. (8): Luminosity of central star.
Col. (9):  Mean Toomre's $Q$ of the disk.
Col. (10): Disk aspect ratio at $r=R_D$.
Col. (11): Measured pitch angle of spiral arms.
Col. (12): Literature references for $i$, $R_D$, $M_*$, the flux densities used to calculate $f_{880\mu \rm m}$, $M_D$, and $L_*$: 
(1) \cite{Benisty2015},
(2) \cite{Andrews2011},
(3) \cite{Dong2018a},
(4) \cite{Muto2012},
(5) \cite{Fairlamb2015},
(6) \cite{Uyama2018},
(7) \cite{vandermarel2016},
(8) \cite{Canovas2018},
(9) \cite{Hashimoto2011},
(10) \cite{Tang2012},
(11) \cite{Andrews2013},
(12) \cite{Pontoppidan2011},
(13) \cite{Boehler2017},
(14) \cite{Ohta2016},
(15) \cite{Kraus2017},
(16) \cite{Benisty2017},
(17) \cite{Menard2019},
(18) \cite{DSHARP3},
(19) \cite{Perez2016},
(20) \cite{DSHARP1},
(21) \cite{Andrews2009},
(22) \cite{Cleeves2016},
(23) \cite{DSHARP4},
(24) \cite{Barenfeld2016},
(25) \cite{Alcala2017}.
}
\end{deluxetable*}

\section{Results and Discussion}

\subsection{Dependence of Pitch Angle on Disk Size, Luminosity, and Mass}

\begin{figure*}
\figurenum{5}
\centering
\includegraphics[width=18cm]{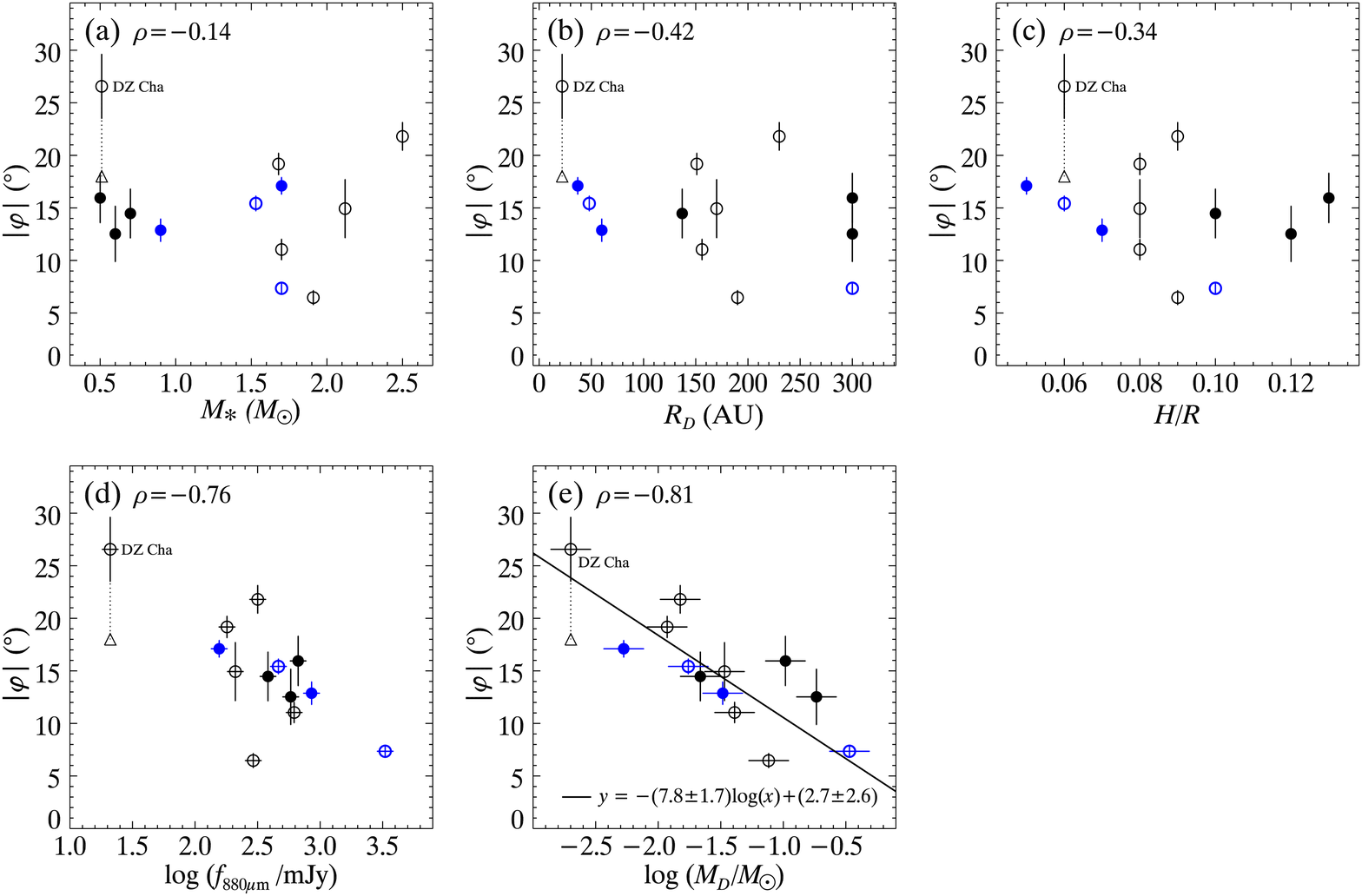}
\caption{Relationship between pitch angle of dominant spiral Fourier mode ($\varphi$) and (a) mass of central 
star ($M_{*}$), (b) disk radius ($R_{D}$), (c) disk aspect ratio ($H/R$), (d) flux density at 880\,$\mu$m 
($f_{880\mu \rm m}$), and (e) disk mass ($M_{D}$).
The open and solid symbols mark the results for NIR scattered-light images and 
ALMA millimeter images, respectively.  The blue symbols denote objects with 
a known companion.  The best-fit function (solid line) in (d) has the form given in 
the bottom of the panel. The Pearson correlation coefficient ($\rho$) is shown in the top of 
each panel.  
The mean pitch angle of the dominant open arm and the weaker tight arm ($18\degr$; triangle) 
of DZ~Cha are especially marked.
}
\end{figure*}

Figure~5 plots the measured pitch angles of dominant spiral Fourier mode against  
the central star mass, disk radius, disk aspect ratio, and disk mass.
The open and solid symbols mark, respectively, 
the results for NIR scattered-light images and ALMA millimeter images, and the 
blue symbols denote the four systems (HD~142527, HD~100453, AS~205, and HT~Lup) with a 
known companion.  The measured pitch angles hardly correlate with the mass of 
the central star (Figure~5a; Pearson correlation coefficient $\rho= -0.14$), but
there is a weak tendency for more tightly wound arms to reside in disks with 
somewhat larger sizes (Figure~5b; $\rho=-0.42$) and higher aspect ratios 
(Figure~5c; $\rho=-0.34$).  
Most strikingly, we found a strong inverse 
correlation between pitch angle and disk mass: smaller pitch angles are 
associated with more massive protoplanetary disks (Figure~5e; $\rho=-0.81$). 
Fitting a logarithmic function gives

\begin{eqnarray}
|\varphi| = -(7.8\pm1.7)\log(M_{D}/M_{\odot})+(2.7\pm2.6),
\end{eqnarray}

\noindent
with a scatter of $3\degr$ in pitch angle or 0.4\,dex in disk mass.  
Since more
massive disks tend to be larger, the weak $\varphi-R_D$ relation is likely a 
secondary manifestation of the stronger primary $\varphi-M_D$ relation. 
Systems with known companions also follow the same empirical trend.  If we only
consider the results from the ALMA millimeter images, the correlation between 
pitch angle and disk mass becomes shallower, but this may be an artifact of the
narrow range of pitch angles ($12\degr-17\degr$) probed by this subset of 
points.  The largest source of uncertainty in Equation (2) lies in the disk 
masses.  Although all the masses were derived using a uniform method 
\citep{Dong2018b}, systematic biases may exist.  In particular, the gas-to-dust
mass ratio adopted in this method has not been well-constrained.  
The estimated disk mass may not be the true mass, but it can be related 
to other physical quantities of the disk (e.g. temperature, size, and dust mass).
Other methods of 
estimating disk mass based on gas tracers (CO or HD) report significantly 
different results \citep[e.g.,][]{Bergin2017}.  
There may be a systematic offset between the absolute disk masses derived from dust emission compared to those derived from
other tracers.  While such a systematic offset will quantitatively change the 
form of the $\varphi-M_D$, the qualitative nature of the physical dependence 
between pitch angle and disk mass should still be preserved.
With these caveats in mind,
Figure~5e suggests that, independent of the formation mechanism, the disk mass 
plays a fundamental role in determining the pitch angle of the observed spiral structure.

In light of the significant uncertainties associated with the disk masses, we verify 
that pitch angle correlates nearly equally well with the model-independent submillimeter 
flux density (Figure 5d; $\rho = -0.76$), such that disks with brighter submillimeter flux density 
tend to have more tightly wound spiral arms.

DZ~Cha deserves special comment.  The pitch angle derived from the dominant spiral Fourier mode 
well traces the stronger arm but not the weaker one, and its high value ($\varphi=26\fdg6\pm3\fdg1$) 
has a strong effect on the empirical trends in Figure 5.  If we exclude DZ~Cha, the $\varphi-R_D$ and 
$\varphi-H/R$ relations become much weaker, with $\rho$ reducing to $-0.22$ and 
$-0.18$, respectively.  But without DZ~Cha the $\varphi-f_{880\mu \rm m}$ ($\rho\prime=-0.56$) and 
$\varphi-M_D$ ($\rho\prime=-0.70$) relations remain strong.
Under the possibility that the mean pitch angle of the two arms of DZ~Cha is more fundamental, setting the pitch angle to $18\degr$ (triangle in Figure~5) reduces the Pearson's correlation coefficient of the $\varphi-f_{880\mu \rm m}$ and 
$\varphi-M_D$ relations to $-0.55$ and $-0.69$, respectively, but still preserves their statistical significance. Therefore, DZ Cha weakens but does not strongly affect the main conclusions of this study.

\subsection{Discussion}

Traditional numerical simulations of isolated protoplanetary disks can generate
transient but recurrent material spiral arms \citep[e.g.,][]{Lodato2004, 
Rice2004, Forgan2011}, but, contrary to observations, they generally produce 
significantly more than two arms.  Although recent simulations show that two 
material arms can also arise \citep{Dong2015b, Tomida2017}, their short life 
times, if owing to gravitational instability, implies that they are statistically less likely to be observed.  The 
frequent change of shape of these material spiral arms makes it difficult to 
maintain pitch angles long-lived enough to produce the observed strong 
$\varphi-M_D$ relation.

Shadowing from a misaligned inner disk can trigger spiral arms detected in 
scattered light \citep{Montesinos2016, Montesinos2018}. Although a less massive disk may be
associated with more open arms at the onset their formation, the arms become tighter 
with time and eventually evolve into tight arms with $\varphi\approx 13\degr$ \citep{Montesinos2016}. 
Besides, the five disks observed in millimeter continuum used in this work do not reveal evidence of a
misaligned inner disk \citep[also see][]{DSHARP3}. 
In particular, there are no signatures of shadowing in IM~Lup in scattered light \citep{Avenhaus2018}.

Studying spiral wakes in a non-gravitating disk excited by a planet, 
\cite{Rafikov2002} showed that the spiral arm pitch angle depends on the 
temperature of the disk and the location of the planet.  Pursuing this further,
\cite{Zhu2015} carried out three-dimensional hydrodynamical simulations and 
found that, 
as a consequence of the non-linear evolution of the spiral wave propagation, 
planet-excited spirals have a larger pitch angle with a more massive perturber.
In this scenario, the pitch angle depends on the location and mass of the 
perturbing plane, not explicitly on the surface density and/or mass of the 
disk.  Thus, this scenario is unlikely to explain our results.

Another possible mechanism may be density wave theory \citep{LinShu64, Bertin1996}.
In a massive thin disk, tightly wound spirals, formed via internal gravitational instability, would have
maximum growth rate at wavenumber $k = \pi G \Sigma/c_{s}^2$, resulting in
pitch angle
\begin{eqnarray}
\varphi \propto \frac{c_s^2}{\Sigma r},
\end{eqnarray}

\noindent
where $c_s$, $\Sigma$, and $r$ are the sound speed, mass surface density, and radial distance \citep{Hozumi2003}.
Adopting a simple power-law distribution $\Sigma \propto \Sigma_0r^{-p}$, where $\Sigma_0 \propto M_D/R_D^2$, and
$c_s^2\propto T\propto L_{\rm *}^{0.25}\,r^{-q}$, we have
\begin{eqnarray}
\varphi \propto \frac{L_*^{0.25}}{M_D}R_D^2\,r^{p-q-1}.
\end{eqnarray}

\noindent
The criterion $Q<1$ for an unstable disk is derived for local axisymmetric Jeans instabilities, but for non-axisymmetric
disturbances the threshold value of $Q$ is marginally larger than 1. Thus, Eq. (4) is valid
for a massive disk with Toomre's $Q$ less than or slightly larger than 1 (Table~1).  This may
explain the trend that pitch angle decreases with larger mass for massive disks, namely the high-mass end of the $\varphi-M_D$ relation. However, light disks are characterized by Toomre's $Q \gg 1$. Equation (4) does not apply to them.  Light disks are stable against gravitational instability, and an external disturber may be to trigger spiral structure.

Even though Eq. (3) is valid only for a massive disk, it may still shed light on understanding spirals in a light disk.
Eq. (3) implies that spirals would be more open (larger $|\varphi|$) if the material 
responding to the perturbation is hotter (higher $c_s$). In other words, spirals observed 
in NIR scattered-light images may be more open than their 
counterparts observed in dust millimeter continuum emission.  
On the other hand, the 3-D structure of spirals can be more complicated than this simple argument.
\cite{JuhaszRosotti2018} found that the spirals
at the disk atmosphere, which is several times hotter than the disk midplane,
are only slightly more open than the spirals at the disk midplane.
MWC~758 was 
observed in both bands.  As the arms of MWC~758 are much less symmetric and 
regular in the dust continuum than in the NIR, we did not attempt to analyze 
its millimeter image.  However, \cite{Dong2018a} show that the spiral arms of MWC~758 
in millimeter continuum indeed are slightly tighter than in NIR scattered light, 
consistent with our expectations.

\section{Summary}

We use two-dimensional Fourier transformation to measure the pitch angle of 
the dominant spiral Fourier mode for 13 protoplanetary disks imaged in 
the NIR in scattered light and in millimeter dust continuum emission.  
The measured pitch angles correlate well with 880 micron flux density, such that disks with brighter 
submillimeter flux densities tend to have more tightly wound spiral arms.
Most strikingly, the pitch angle exhibits 
a strong inverse correlation with the disk mass, following 
$|\varphi| = -(7.8\pm1.7)\log(M_{D}/M_{\odot})+(2.7\pm2.6)$.
Four disks with a known 
companion also obey this scaling relation.  Such a strong dependence of pitch 
angle on disk mass is not expected in the theory or hydrodynamical simulations 
of non-gravitating disks.  
In contrast, density wave theory may partly explain the $\varphi-M_D$ relation in the high-mass end.
Our result suggests that disk mass, independent of the formation mechanism, 
plays a fundamental role in determining the pitch angle of the observed spiral arms.  
The empirical correlation revealed in this work provides a simple empirical, independent 
method to use the pitch angle of spiral arms to constrain the mass of protoplanetary disks.

\begin{acknowledgements}
SY and LH acknowledge support from the National Science Foundation of China (11721303) and the National Key R\&D Program of China (2016YFA0400702).  ZZ acknowledges support from the National Science Foundation under CAREER Grant Number AST-1753168 and the Sloan Foundation.  We thank the very constructive suggestions from the referee.
We thank Ruobing Dong, Gregory Herczeg, and Feng Long for helpful discussions 
and valuable advice.  We are grateful to Henning Avenhaus, Myriam Benisty, 
Hector Canovas, Antonio Garufi, and Jun Hashimoto for making available the 
observational images used in Figures 1 and 2.  This paper makes use of the 
following ALMA data: \\ ADS/JAO.ALMA\#2016.1.00484.L.  ALMA is a partnership 
of ESO (representing its member states), NSF (USA), and NINS (Japan), together 
with NRC (Canada), NSC and ASIAA (Taiwan), and KASI (Republic of Korea), in 
cooperation with the Republic of Chile. The Joint ALMA Observatory is operated 
by ESO, AUI/NRAO, and NAOJ.
\end{acknowledgements}

\end{document}